# A review of recent theoretical and computational studies on pinned surface nanobubbles

Yawei Liu (刘亚伟), Xianren Zhang (张现仁)[†]

*State Key Laboratory of Organic−Inorganic Composites, Beijing University of Chemical Technology, Beijing 100029, China*

The findings of long-lived surface nanobubbles in various experiments brought a puzzle in theory, as they were supposed to be dissolved in microseconds due to the high Laplace pressure. However, an increasing number of studies suggest that the pinning of contact line, together with certain levels of oversaturation, is responsible for the anomalous stability of surface nanobubble. This mechanism can interpret most characteristics of surface nanobubbles. Here we summarize recent theoretical and computational work to explain how the surface nanobubbles become stable with the pinning of contact line. Other related work devoted to understand the unusual behaviors of pinned surface nanobubbles are also reviewed here.

[†] Corresponding author. E-mail: zhangxr@mail.buct.edu.cn

# 1. Introduction

It has been nearly 20 years since the first images of surface nanobubbles were reported by two independent groups working in China[1] and Japan[2]. Actually, the first clue of the existence of such surface nanobubbles can be retraced from the earlier surface force measurements between hydrophobic surfaces immersed in aqueous solutions[3–5]. As early as 1994, Parker, Claesson and Attard proposed that the stepwise features in force curves from their experiments were due to the bridging of nanoscale bubbles adhered on solid surfaces[3]. Following from these initial reports of surface nanobubbles, many experimental researches employed different techniques such as AFM measurement[1,2,6–15], rapid cryofixation[16], neutron reflectometry[17], and direct optical visualization[18,19] have shown that the surface nanobubbles can exist for a substantially long period of time.

Surface nanobubbles are of great interest as they have great potential applications in many fields, such as boundary slip in fluid[20,21], froth-flotation[22], and protein adsorption[23,24]. However, the stability of surface nanobubbles brought a conundrum for theory[25–27]. Theoretically, bubbles at the nanoscale are unstable due to the sharply increase of the Laplace pressure. The Laplace pressure, $\Delta p$, for a spherical bubble of radius $R$ is given by $\Delta p = 2\gamma/R$ with $\gamma$ the surface tension of the bubble interface. Therefore, smaller bubbles have higher internal pressure. For example, the extra internal pressure is $\sim 14.3$ atmospheres for a bubble of radius $100\ nm$ in water. The increased internal pressure leads to an increase of the solubility of the gas in the solution surrounding the bubble. Consequently, if the solution around the bubble is not sufficiently oversaturated, gas will leave the bubble by diffusion in order to establish equilibrium. The loss of gas from the bubble leads to a decrease in the bubble size, which further raises the Laplace pressure and thus results in more gas leaving the bubble. As a result, a positive feedback cycle is established and rapidly lead to the dissolution and disappearance of small bubbles. The Epstein and Plesset theory[28], which was developed to describe the gas diffusion process around a bubble, predicts that in saturated solutions small bubbles should very rapidly shrink and disappear. For example, the lifetime of bubbles smaller than 1000 nm is less than 0.02 s, making it is difficult to detect and measure such bubbles in experiments[29]. Note that if the solution is oversaturated with gas, the positive feedback cycle will operate in reverse and will result in the rapid growth of bubbles from small bubbles, which also has been discussed by Epstein and Plesset[28]. In short, the earlier studies based on classical theories predicted that there is no thermodynamic stability for surface nanobubbles, and such bubbles should shrink and disappear or grow to a macroscopic size in a very short time.

Different mechanisms, including the dynamic equilibrium theory[30,31] and the contamination theory[32], have been proposed and provided important insights, but both of them are insufficient to account for stable surface nanobubbles[33]. In 2012, we proposed that pinning of contact line induced by surface heterogeneities (e.g. physical roughness or chemical heterogeneities) leads to the appearance of thermodynamically metastable surface nanobubbles, which can explain the existence of long-lived surface nanobubbles in experiments[34]. The contribution of contact line pinning to the nanobubble stability has also been proposed in the theoretical work by Weijs and Lohse[35], and in the experimental work by Zhang *et al.*[36]. Several recent experimental work on surface nanobubbles also supported that the pinning of contact line is a necessary condition for achieving the stability of surface nanobubbles[37–41].

An impressive amount of experimental, theoretical, and computational work on surface nanobubble has been accumulated over the last 17 years. Here we do not attempt to discuss the entire corpus of this field, for which the readers are encouraged to consult the more comprehensive reviews[25–27,29,33,42–44]. Rather, we focus on recent theoretical and computational studies, especially those from our group, of pinned surface nanobubbles. From Sec. 2 to Sec. 5, we review related researches on the pinned surface nanobubbles with different theories and methods, including the classical nucleation theory (CNT), the Epstein and Plesset theory, the lattice density functional theory (LDFT) and the molecular dynamics (MD) simulation method. Finally, in Sec. 6 we put forward conclusions and give a perspective on the present and future challenges.

## 2. Classical nucleation theory

On the basis of the pioneering work of Gibbs, Becker and Doring[45], Volmer and Weber[46] and others developed the classical nucleation theory (CNT). According to the CNT, the free energy cost for the formation of a spherical bubble in the bulk liquid consists of a volume term and a surface term: $\Delta G = -(4\pi/3)R^3 \cdot \Delta p + 4\pi R^2 \cdot \gamma$ where $\Delta p = p_v - p_l$ is the pressure difference between the center of bubble and the bulk liquid, $\gamma$ is the surface tension and $R$ is the radius of the bubble. By setting $\partial \Delta G/\partial R = 0$, the free energy reaches its maximum at a critical radius $R^* = 2\gamma/\Delta p$, which has the same formula as the Laplace equation for the mechanical stability of a bubble. Thus, bubbles will grow infinitely if their size exceeds the critical size, or they will disappear if they have a size less than the critical size.

When there exist surface heterogeneities that can pin the contact line of surface nanobubbles, the situation changes. In our previous work[34], we found that within the framework of CNT, a stable surface nanobubble corresponds to a free energy minimum when the contact line is pinned. The same conclusion was obtained by the subsequent theoretical work by Attard[47], in which the author used the modifying classical nucleation theory to include the oversaturation dependence of the surface tension.

CNT suggests that for stable surface nanobubbles, a relationship of $\sin\theta = r/R$ with $\theta$ the contact angle of the nanobubble, $r$ the footprint radius of the bubble, and $R$ the surface radius that equals to the critical radius (i.e. $R = R^*$), holds. The critical radius $R^*$ can be further written as $R^* = 2\gamma/\Delta p = 2\gamma/(\rho_v \Delta\mu)$ with $\rho_v$ the density of vapor and $\Delta\mu$ is the difference of chemical potential between liquid and vapor that determines the level of oversaturation. Therefore, this relationship predicts how the contact angle depends on the oversaturation and the footprint radius, as well as the upper threshold of the footprint radius (i.e. $r < R^*$).

Within the framework of CNT, a negative feedback mechanism was proposed to explain the stability of pinned surface nanobubbles[34,48]. The Laplace equation for a bubble in equilibrium with its surrounding, $\Delta p = 2\gamma/R$, describes the mechanical balance on the interface between the contribution from $\Delta p$, which makes the bubble grow, and that from $\gamma$, which makes the bubble shrink. Consider a pinned surface nanobubble as shown in Fig.1 (b), and a small perturbation of the bubble's volume. If the volume of the nanobubble increases, the curvature radius of the bubble would decrease because the contact line cannot move, and therefore the contribution of $2\gamma/R$ increases. In this case, $\Delta p < 2\gamma/R$, so that the bubble shrinks accordingly and returns to the equilibrium state. If, conversely, the nanobubble shrinks in its volume under a small perturbation, its curvature radius would increase, and thus $\Delta p > 2\gamma/R$, leading to the growth of the bubble. Therefore, a negative feedback cycle is established to prevents the nanobubble from shrinking and growing, and stabilizes the nanobubble. This negative feedback mechanism only occurs when the contact angle of nanobubbles (measured in the liquid phase) is larger than $90°$, which can explain that surface nanobubbles always have increased contact angle compared with the corresponding angle of a macroscopic bubble or droplet on the same surface[2,8,26,33,49]. On the other hand, if there is no pinning of contact line, the equilibrium bubble is not stable and will rapidly disappear or grow under the positive feedback mechanism [see Fig.1 (a)].

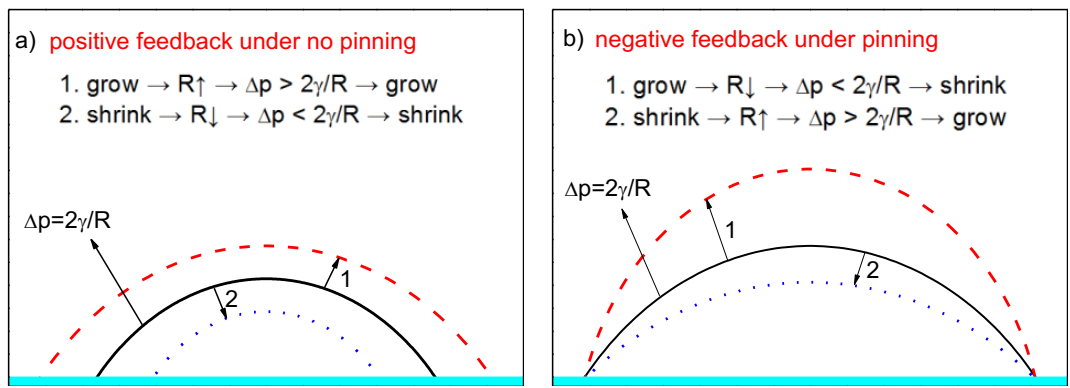

**Fig.1** Schematic illustration of the positive feedback mechanism for unstable nanobubbles without pinning of contact line pinning (a) and that of the negative feedback mechanism for stable nanobubbles with pinning of contact line pinning (b). The figure is reproduced from Ref. [48].

**3. Epstein and Plesset theory**

The kinetics of the gas diffusion process around a bubble has been dealt in 1950 by Epstein and Plesset[28]. A gas bubble in a solution will grow or shrink by diffusion, depending on the solution being oversaturated or undersaturated. To solve the diffusion equation, the local concentration of gas dissolved in the solution adjacent to bubble surface is determined by applying Henry's law, at which the total pressure is the sum of the external pressure and the Laplace pressure. Therefore, the evolution of the bubble can be determined by diffusion equation, Laplace equation, and Henry's law.

For oversaturation (the gas oversaturation $\xi > 0$), Lohse and Zhang[50] provided an exact calculation for a single pinned surface nanobubble and confirmed that the bubble can reach a stable equilibrium state, for which the bubble satisfies the relationship $\sin\theta = \xi L/L_c$ with $L$ the bubble's footprint diameter and $L_c$ the critical footprint diameter. This relationship can return into the one obtained from CNT (see Sec. 2) by using $L = 2r$, $L_c = 4\gamma/p_0$ ($p_0$ the ambient pressure), $\Delta p = \xi p_0$ (Henry's law) and $R^* = 2\gamma/\Delta p$ (the critical radius in CNT). Therefore, above relationship also predicts the contact angle according to the oversaturation and the footprint diameter, as well as the upper threshold of the footprint diameter (i.e. $L < L_c/\xi$).

For undersaturation ($\xi < 0$), the theoretical work from Weijs and Lohse[35] confirmed that pinned surface nanobubbles cannot reach a stable equilibrium state, but dissolve on a much longer timescale than free bubbles, which can explain why under normal experimental conditions (e.g. the liquid is exposed to atmospheric conditions) surface nanobubbles can live for many hours or even up to days.

**4. Lattice density functional theory**

The lattice density functional theory (LDFT)[51–57] based on a simple cubic lattice gas model provides a very simple but efficient method to investigate the behavior of simple fluids at a molecular level, and has been widely used to study the capillary condensation and evaporation[51–55], the vapor-liquid nucleation[58–62], the water bridges[63] and the wetting of solid surface[64,65]. In these years, LDFT has also been used to study the pinned surface nanobubbles[34,66–70].

**4.1. Thermodynamically metastable pinned surface nanobubbles**

LDFT was first used to explain the stability of surface nanobubbles in 2012[34]. In that work, the free energy changes as a function of nanobubble's volume on various substrates were calculated by using the constrained LDFT[56]. It was found that for oversaturated liquid on substrates having sufficient surface heterogeneities, surface nanobubbles can be in a thermodynamically metastable state: the state with nanobubble corresponds to a local minimum of free energy. The kinetic LDFT[52] calculations showed that if the pinning effect is absent, the surface nanobubble would dissolve with roughly constant contact angle. But if there existed the pinning effect, both the volume and the contact angle of the surface nanobubble remained constant. Examples of stable pinned surface nanobubbles in oversaturated liquids, obtained by LDFT, are shown in Fig. 2.

The LDFT calculations also revealed that stable pinned surface nanobubbles always have a contact angle (measured from the liquid side) larger than 90°. Importantly, the contact angle was found to depend on both the oversaturation and the footprint radius of nanobubbles, but is independent on the substrate wettability [34]. These results agree well with the theoretical results (see Sec. 2 and 3) and experimental observations that the contact angle of nanobubbles is size dependent and substrate wettability independent[2,8,26,33].

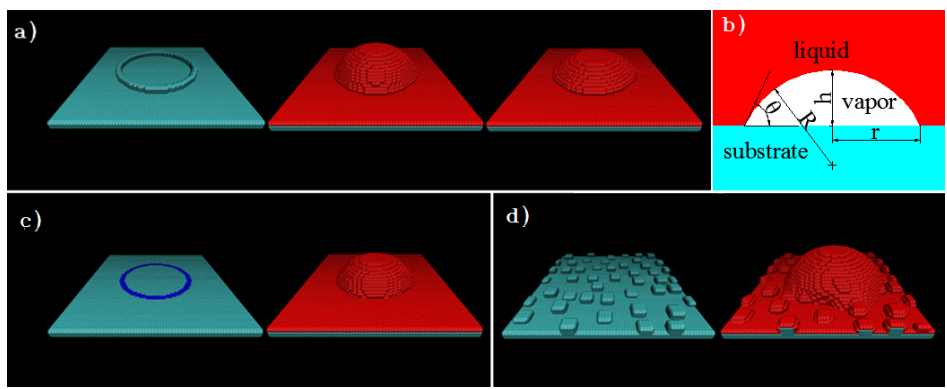

**Fig.2** Stable surface nanobubble induced by the contact line pinning effect. a) From left to right, the substrate with a ring pattern and stable surface nanobubbles at two different degrees of supersaturation. b) Schematic drawing of the system. c) The substrate with the chemical heterogeneity in a ring-like shape and the stable surface nanobubble. d) The solid surface with randomly distributed physical heterogeneity and the stable

surface nanobubble. The figure is reproduced from Ref. [34].

**4.2. Relationship between substrates and pinned surface nanobubbles**

LDFT was then used to investigate how the nanobubble stability depends on the substrate structure (roughness) and wettability[66]. In the calculations four types of substrates with different local structures (see Fig.3) were considered: Type I (circular terrace), Type II (circular concave), Type III (ring), and Type IV (perforated ring) substrates. The substrate wettability was controlled by adjusting the strength of fluid-solid interaction, $\varepsilon_{sf}$. Fig.3 also shows typical surface nanobubbles stabilized on those substrates at given oversaturation and pinning radius (i.e. bubble's footprint radius). Therefore, both the substrate roughness and wettability affect the nanobubble stability. However, those stable surface nanobubbles show the same morphology, indicating that the contact angle of pinned surface nanobubble is independent on the substrate wettability, which again agrees with the predictions from theories (see Sec. 2 and 3).

To explain those results from LDFT calculations, the pinning force, $f_p$, which acts on the contact line of a nanobubble and prevents its lateral motion, was quantitatively evaluated according to the corresponding definition: $f_p = \gamma(\cos\theta_{bubble} - \cos\theta_0)$ with $\gamma$ the surface tension, $\theta_{bubble}$ the contact angle of the nanobubble, and $\theta_0$ the contact angle corresponding to a bubble on smooth substrates[66]. Both the pinning force to stabilize nanobubbles and that the substrate can provide were calculated. It was found that the substrate structure and wettability together determine the sign and threshold of the pinning force provided by the substrate, whereas the pinning force required to stabilize the nanobubble is related to the substrate wettability and the pinning radius. If and only if the required pinning force is within the range of the provided pinning forces, contact line pinning occurs and therefore the nanobubble is stable. Otherwise, the motion of the contact line occurs and the nanobubble becomes unstable.

Those calculations revealed the relationship between nanobubbles and substrates: "the substrate structure and wettability determine the possible range of pinning force and thus affect the nanobubble stability. However, for stable nanobubble, the contact angle is independent of substrate wettability."

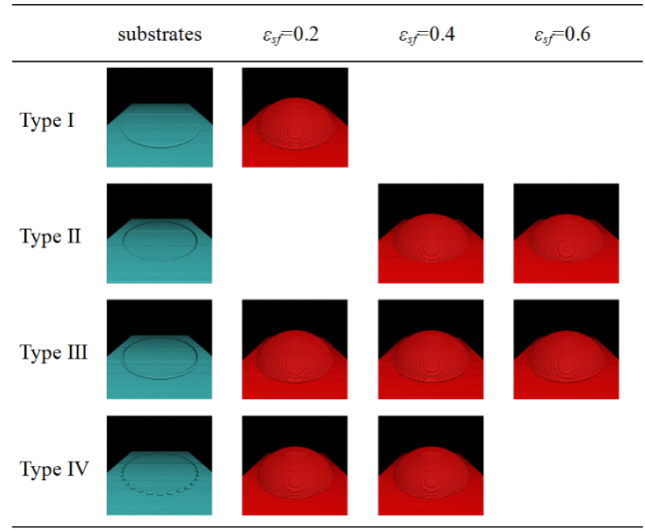

**Fig.3** Snapshots of four types of substrates (the first column) and stable surface nanobubbles (the second to fourth columns) on the substrates. From left to right, the substrate's wettability is changing by adjusting the fluid-solid interaction strength, $\epsilon_{sf}$. The figure is reproduced from Ref. [66].

**4.3. Interaction between AFM tips and pinned surface nanobubbles**

In most experiments, the morphology of surface nanobubbles was imaged with AFM[1,2,6–15]. The interaction between AFM tips and the surface nanobubbles certainly will influence the final images of such bubbles. LDFT was also adopted to explore the process when the AFM tip approaching toward, contacting with, and finally retracting from the pinned surface nanobubbles, and to unravel the interaction between them, as well as how the wettability and shape of the AFM tip affect the morphology of surface nanobubbles[69].

In their calculations, the system contained an AFM tip and a stable pinned surface nanobubble on a rough substrate. The rough substrate was modeled as a solid surface decorated by a ring-shaped pattern with a radius of $20\sigma$ ($\sigma$ is the lattice spacing), and the AFM tip was modeled as a hemispherical end on top of a solid cylinder with a radius of $10\sigma$. A cone-shaped tip was also built to investigate the effect of tip shape.

LDFT calculations [69] indicated that the nanobubble showed an elastic deformation for the approach of a hydrophilic tip [Fig.4 (a)]. The hydrophilic nature of the tip and a thin wetting film covering it prevented the tip from penetrating the bubble during an approach process, showing an elastic effect. It was the elastic effect that induces strong nanobubble deformation and repulsive interaction. In the retraction process, on the other hand, the hydrophilic tip could immerse the bulk solution easily through departure from the nanobubble surface, leading to a weak and intermediate-range attraction between the tip and the nanobubble. For hydrophobic tip, however, a different situation took place [see Fig.4 (b)]. In this case, the nanobubble shoeds an adhering effect, which dominates the tip-nanobubble interaction. In the approach process, the hydrophobic nature of the tip facilitated its penetration through the vapor−liquid interface without causing significant nanobubble deformation, which results in the disappearance of the repulsive force. In retraction process, on the other hand, the hydrophobic adhesion between the tip and the nanobubble induced a much strong lengthening effect on the nanobubble deformation. As a result, a strong and long-ranged attractive force was observed.

On the other hand, LDFT calculations showed that a cone-shaped tip also induced the deformation of the nanobubble during the approaching and retracting process, however, in a much weaker manner. Therefore, a sharp-ended hydrophilic tip maybe a good choice to design minimally invasive experiments, which was in good agreement with the conclusion in Walczyk and Schönherr's experimental work[71].

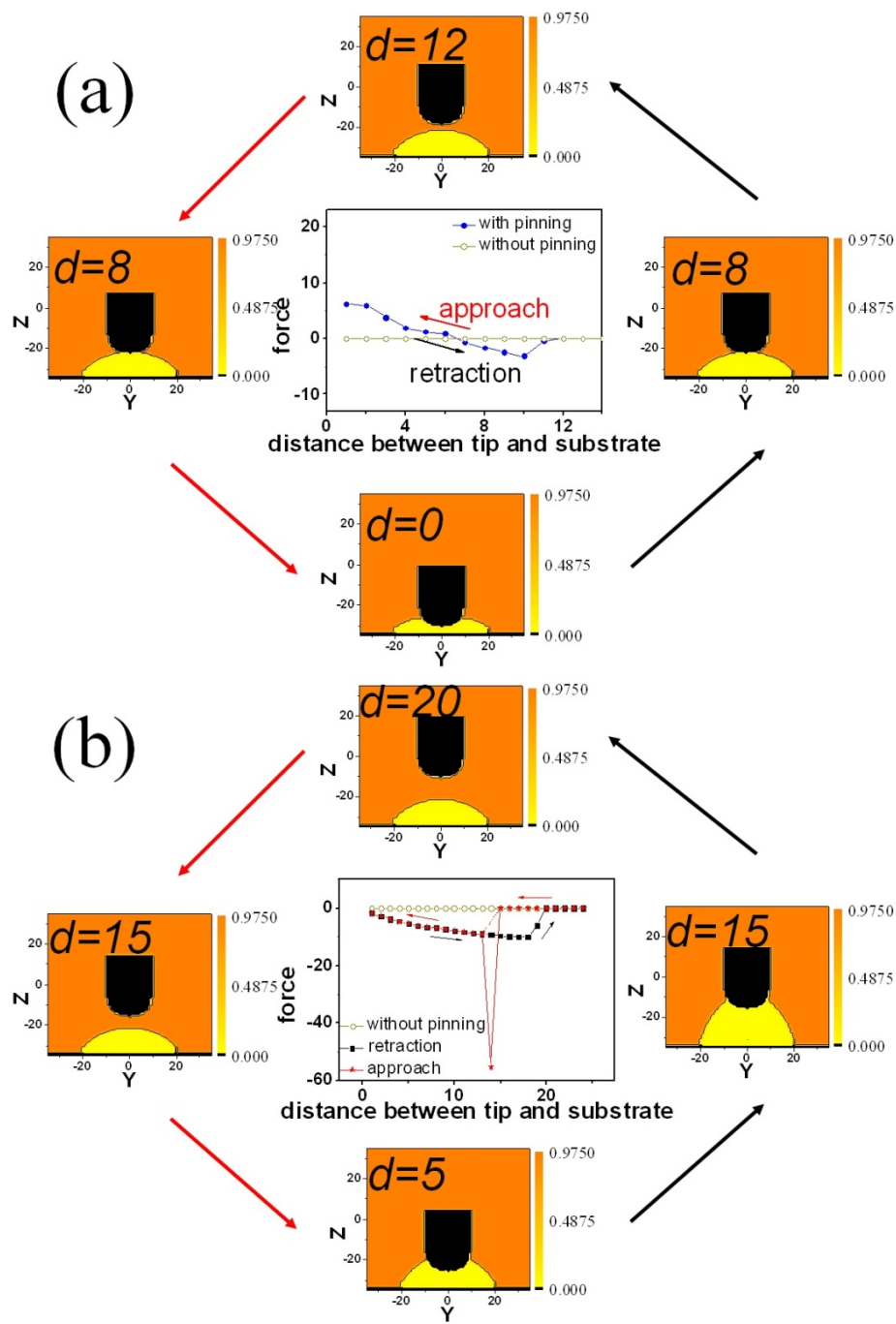

**Fig.4** Capillary force between AFM tip and nanobubble and the morphology of the nanobubbles for different tip-substrate distances in the approach (red arrow) and retraction process (black arrow): (a) for a hydrophilic tip and (b) for a hydrophobic tip. The figure is reproduced from Ref. [69].

## 5. Molecular dynamic simulations

Compared to LDFT calculations, MD simulations are increasingly used in investigating surface nanobubbles[72–76].

**5.1 Pinning and oversaturation induce stable surface nanobubbles**

Stable pinned surface nanobubbles were first achieved in MD simulation in 2014[72]. The stability of surface nanobubbles in both pure fluids and gas-liquid mixtures was studied. In the computational work, a fluid system with LJ potentials to describe the interaction between the different species was employed. The LAMMPS package[77] was used, with constant temperature, pressure, and particle number. Note that in the MD simulations, the pressure was controlled by applying an external force on the wall at the top of simulation box. A nanopore in the bottom substrate was explicitly included into the substrate to induce the pinning of contact line. In the case with gas-liquid mixtures, a reservoir of gas molecules, in which the identity exchange of liquid and gas molecules in the reservoir was performed every 0.1 ns, was introduced to maintain a target gas concentration far from the surface nanobubble.

In the MD simulations[72], stable pinned surface nanobubbles were found in both pure liquids and gas-liquid mixtures, provided that there was suitable overheating or gas oversaturation [Fig.5 (a) and (c)]. A Wenzel or a Cassie wetting state was found in undersaturated and saturated fluids, and the liquid-to-vapor phase transition occurred at high oversaturation [Fig.5 (b) and (d)]. Besides, MD simulations proved that an excess of gas molecules dissolved is required to produce stable gas nanobubbles, which can explain why no nanobubble is observed in degassed water at room temperature in experiments. However, for surface nanobubbles either in pure liquids or in gas-liquid mixtures, both the curvature radius and the contact angle of nanobubbles increase as the level of overheating/oversaturation decreases, which again agrees with the predictions from theories (see Sec. 2 and 3). These MD simulations of pinned surface nanobubble[72] unambiguously showed that the contact line pinning effect, which together with the oversaturation turned out to be crucial for surface nanobubble stabilization.

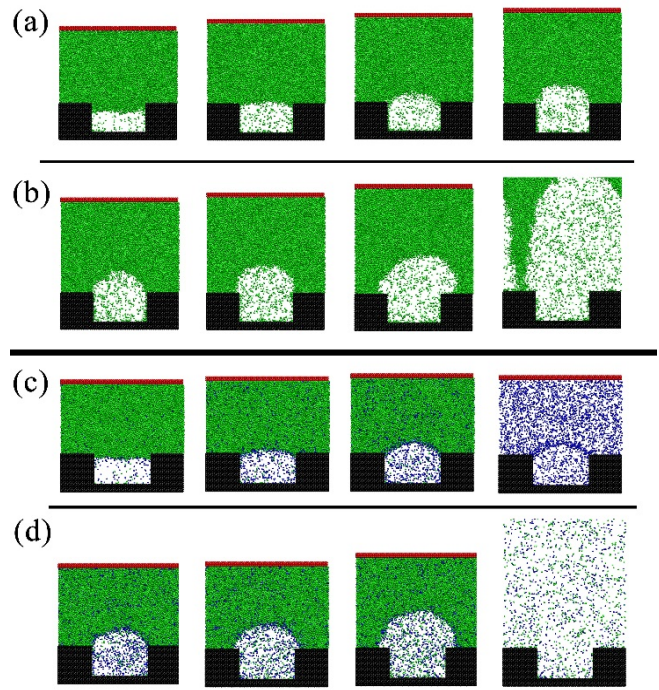

**Fig.5** (a) Stable pinned surface nanobubble in pure liquids at different degrees of overheating. From left to right, the degree of overheating is increasing. (b) Several typical snapshots during the liquid-to-vapor phase transition in pure liquids process at a high degree of overheating. (c) Stable pinned surface nanobubble in gas-liquid mixtures at different degrees of oversaturation. From left to right, the degree of oversaturation is increasing. (d) Several typical snapshots during the liquid-to-vapor phase transition process in gas-liquid mixtures at a high degree of oversaturation. The figure is reproduced from Ref. [72].

**5.2 Nucleation of pinned surface nanobubble on rough substrates**

MD simulations were then used to study how nanobubbles form on rough hydrophobic substrates[78]. The simulation system employed in that work was similar to that in Ref. [72], except that the rough substrates were modeled as solid surfaces decorated with several identical nanopillars. Three different numbers of nanopillars were used to study the nanobubble nucleation at substrates having high, moderate, and low degree of surface roughness. The substrates were hydrophobic, and the gas-liquid mixture was set to be oversaturated by controlling the gas concentration in a gas reservoir.

First, long-time standard simulations were carried out to directly observe the kinetic pathways[78]. A two-step nucleation route involving the formation of an intermediate state was found for the nanobubble formation: at first, several gas cavities occurs in the grooves, leading to a Wenzel-to-Cassie transition [e.g. (i)-(iv) in Fig.6(b)]; then, those gas cavities coalesce and form a stable pinned surface nanobubble, inducing a Cassie-to-nanobubble transition [e.g. (iv)-(v) in Fig.6(b)]. Additionally, the corresponding free energy changes were quantitatively evaluated[78] by using constrained simulations combined with the thermodynamic integration approach. The free energy barriers for the two sequential transitions have opposing dependencies on the degree of surface roughness, so that surface nanobubbles are more likely to form on the surfaces with moderate roughness as each state transition only need to overcome a moderate free energy barrier.

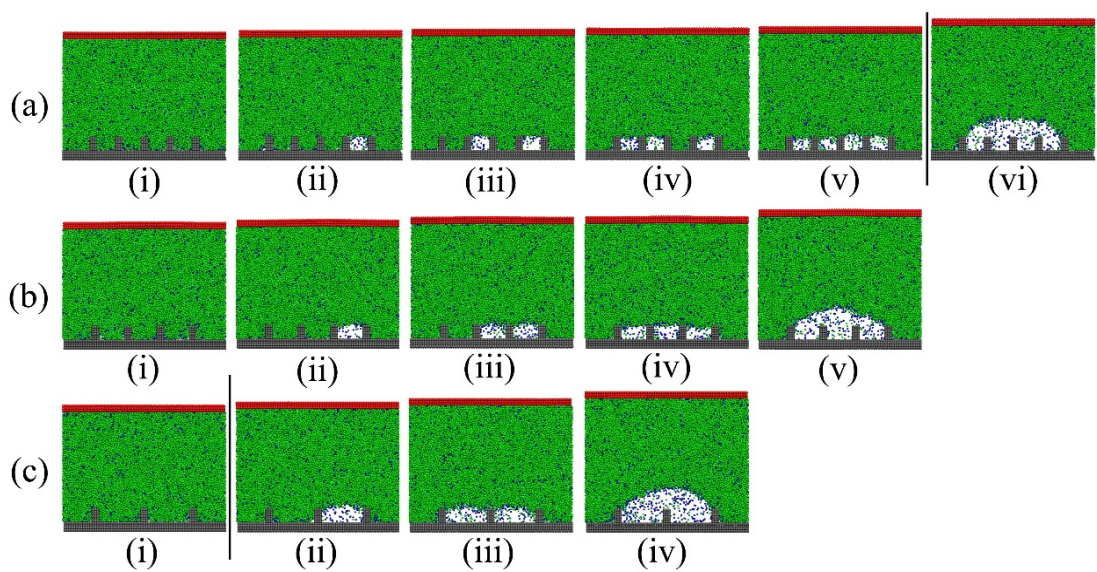

**Fig.6** Several typical snapshots during the nanobubble formations at surfaces with high roughness (a), moderate roughness (b), and low roughness(c). Snapshots (i)-(v) in (a), (i)-(v) in (b), and (i) in (c) are obtained from the standard long-time MD simulations while others are obtained from the constrained MD simulations using the thermodynamic integration approach. Reprinted with permission from Ref. [72].

**5.3 Solvent exchange producing pinned surface nanobubbles**
In experiments, the oversaturation environment that is required to the stable pinned surface nanobubbles is often produced through the solvent exchange procedure. For example, Lou *et al.*[1] used water (poor solvent for air) to exchange the ethanol (good solvent for air) to produce surface nanobubbles. MD simulations recently was used to understand molecular mechanisms behind the solvent exchange[76]. The simulation system was similar to that in Ref. [72]. The solvent exchange was achieved as follow: initially, the box was full of a good solvent; then, during the simulations, the identity exchange between liquid and gas molecules as well as the solvent exchange was performed at the same time interval, to keep stable gas concentration and solvent composition in a source region that was far from the substrate.

MD simulations demonstrated a two-stage mechanism for forming nanobubbles through a process of solvent exchange[76]. During the first stage of the process, an interface between two interchanging solvents was found, which moves toward the substrate gradually as the exchange process proceeds. Unexpectedly, driven by the solubility gradient of liquid composition across the moving solvent−solvent interface, the directed diffusion of gas molecules against gas concentration gradient was found. The forced diffusion against the gas density gradient prevents the gas molecules from washing away, and more importantly it produces an increasingly high local gas oversaturation as the interface approaches the substrates. At the second stage, the local high gas oversaturation nucleates nanobubbles either on the solid surfaces or in the bulk solution, depending on the hydrophobicity of the substrate [see Fig.7].

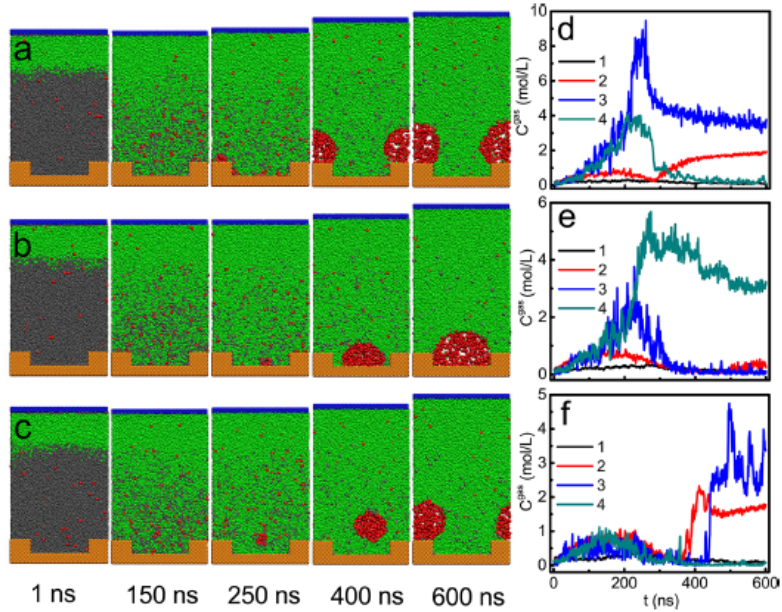

**Fig.7** (a−c) Typical snapshots at different simulation time and (d−f) time evolution of the local gas density in different regions during the solvent exchange processes. The Young's contact angle for substrates are 130° (a, d), 91° (b, e) and 31° (c, f), respectively.

## 6. Conclusions

The origin of the stability of surface nanobubbles is a controversial topic for a long time. Now, an increasing number of experimental, theoretical, and computational work confirmed that the pinning of contact line, which results from the intrinsic nanoscale physical roughness or chemical heterogeneities of substrates, is crucial for the nanobubble stability.

Contact line pinning implies there exists the negative feedback mechanism to stabilize surface nanobubbles: for increasing bubble volume, the Laplace pressure that makes the bubble shrinks increases accordingly, very different from free nanobubbles, for which the Laplace pressure decreases once the bubble volume increases. Under the pinning of contact line, both the classical nucleation theory and the classical Epstein and Plesset theory[28] predicted that surface nanobubbles can reach a thermodynamically stable state in oversaturated liquids/solutions[34,50], following a simple size constraint, i.e. $\sin\theta = r/R$ and $R = R^* = 2\gamma/\Delta p = 2\gamma/(\rho_v \Delta\mu)$. This size constraint shows that the contact angle only depends on the oversaturation and the footprint radius of the nanobubble, and can explain the anomalous contact angle for surface nanobubbles observed in experiments. These conclusions were further confirmed by using the lattice density functional theory (LDFT)[34] and molecular dynamics (MD) simulations[72].

Following from these discussions for the stability of pinned surface nanobubble, both LDFT and MD simulations were widely utilized to reveal various behaviors of pinned surface nanobubbles, such as the pinning force[66], the interaction between AFM tips and nanobubbles[69], the nanobubble formation[76,78].

Although those studies give us useful insights for pinned surface nanobubbles, several challenges and limitations in these calculations and simulations still persist. The lattice nature embedded in LDFT calculations limits its ability to provide more molecular details. The single-component model employed in present LDFT calculations fails to understand the role of dissolved gas in surface nanobubbles. Although stable surface nanobubbles exist in both pure liquid and gas-liquid mixtures[72], dissolved gas was found to have significant influence on bubble growth dynamics[79]. Thus, it would be of interest to use the LDFT extended to binary mixture model[54] in future researches. The effect of dissolved gas can be easily included in MD simulations. However, due to the limit of compute resources, in most present MD simulations of pinned surface nanobubbles, only LJ potentials have been used, with a length scale of at most tens of nanometers, and a time scale of at most hundreds of nanoseconds. It would also be of interest to use more real models and perform longer simulation to gain simulation data that could be directly compared with experimental observations. Of course, other computational method should be considered to applied in this field. For instance, the lattice Boltzmann method[80], which is an approach in between continuum dynamics simulations and MD simulations, may have great potential to study the collective effect[81] in the dissolving and formation process of surface nanobubbles[33]. This kind of studies would certainly deepen our understanding on cavitation and bubble nucleation and collapse[82-87].

**Acknowledgment**

This work is supported by National Natural Science Foundation of China (No. 91434204).